\documentclass[a4paper,10pt]{revtex4}
\usepackage{graphicx} 
\usepackage{amsmath} 
\usepackage{amssymb} 
\usepackage{bm} 
\usepackage{dcolumn}
\usepackage{color}
\usepackage{mathrsfs}
\usepackage{amsfonts}
\usepackage{varioref}
\usepackage{mathrsfs}
\usepackage{graphicx}
\usepackage{latexsym}
\usepackage{amsmath}
\usepackage{amssymb}
\usepackage{textcomp}
\usepackage{amsbsy}
\usepackage{graphics}
\usepackage{epstopdf}
\usepackage{color}
\usepackage[caption=false]{subfig}

\RequirePackage[colorlinks,citecolor=blue,urlcolor=magenta,linkcolor=blue]{hyperref}
\input epsf

\allowdisplaybreaks[4]
\begin{document}

\tolerance=5000

\title{ACT inflation and its influence on reheating era in Einstein-Gauss-Bonnet gravity}

\author{Sergei~D.~Odintsov$^{1,2}$\,\thanks{odintsov@ieec.uab.es},
Tanmoy~Paul$^{3}$\,\thanks{tanmoy.paul@visva-bharai.ac.in}} \affiliation{
$^{1)}$ ICREA, Passeig Luis Companys, 23, 08010 Barcelona, Spain\\
$^{2)}$ Institute of Space Sciences (ICE, CSIC) C. Can Magrans s/n, 08193 Barcelona, Spain\\
$^{3)}$ Visva-Bharati University, Department of Physics, Santiniketan - 731 235, India.}


\tolerance=5000

\begin{abstract}
We investigate the observational viability of non-minimally coupled scalar-Einstein-Gauss-Bonnet (GB) gravity, during inflation and post-inflationary reheating dynamics, from the perspective of the latest ACT-DR6 combined with the Planck 2018 and BAO data. It turns out that the ACT result considerably affects the inflationary e-fold number compared to the case where only Planck 2018 data is taken into account. The viable parameter spaces corresponding to the inflationary ACT-DR6+Planck18+BAO substantially influence the reheating phenomenology via the reheating equation of state ($w_\mathrm{eff}$) and the reheating temperature. In particular, the ACT-DR6+Planck18+BAO data seems to disfavor $w_\mathrm{eff} < 1/3$ during the reheating stage, which is unlike to that of only Planck 2018 case. These reveal how the ACT-DR6 data hits the early universe phenomenology from inflation to reheating in the context of higher curvature like scalar-Einstein-GB theory of gravity.
\end{abstract}

\maketitle

\section{Introduction}\label{SecI}
Currently we are living in a cosmological era which has witnessed a giant leap of understanding on early universe phenomenon, primarily from Cosmic Microwave Background (CMB) \cite{CMB-S4:2016ple,Planck:2018jri}. The recent Atacama Cosmology Telescope (ACT) data, combined with the Planck 2018 and the Baryon Accoustoc Oscillation (BAO) from DESI, constrain the inflationary models in a way that the curvature perturbation should be less red tilted \cite{ACT:2025fju,ACT:2025tim}. In particular, the Data Release 6 (DR6) from the ACT, combined with the Planck 2018 and BAO (collectively it can be denoted by ACT-DR6+Planck18+BAO), refines the constraint on inflationary observables (like the spectral index of primordial curvature perturbation and tensor-to-scalar ratio) as: $n_s = 0.9743 \pm 0.0034$ and $r < 0.038$ respectively \cite{ACT:2025fju,ACT:2025tim}, indicating a $2\sigma$ deviation from the only-Planck 2018 measurements \cite{Planck:2018jri}. Such a refinement on $n_s$ and $r$ from ACT data disfavors the models which previously lie at lower boundary value of $n_s$, however allows such models which produce higher value of $n_s$ such as $\alpha$-attractor inflationary model. Based on these argument, a variety of inflationary models (which are viable from only-Planck 2018 result) are revisited from the angle of the recent ACT data \cite{Kallosh:2025rni,Aoki:2025wld,Odintsov:2025wai,Haque:2025uga,Dioguardi:2025vci,Salvio:2025izr,Antoniadis:2025pfa,Kim:2025dyi,Gao:2025onc,Drees:2025ngb,Maity:2025czp,Mondal:2025kur,Peng:2025bws,Liu:2025qca}.

Beside the inflationary phenomenology, the new constraint on inflationary observables coming from the ACT data also substantially influences the post-inflation reheating dynamics through the reheating equation of state (EoS) parameter ($w_\mathrm{eff}$). For instance, the authors of \cite{Haque:2025uri} recently showed that the $\alpha$-attractor model with T-type potential (i.e. the tan-hyperbolic type of scalar potential) is viable with the ACT-DR6+Planck18+BAO data only when the reheating EoS parameter is larger than a certain value given by $w_\mathrm{eff} \gtrsim 0.44$, which is unlike to that of the case when only Planck 2018 measurement is taken into account. The important point to understand is that a change on the viable regime of $w_\mathrm{eff}$ directly affects the evolution of the primordial gravitational waves (GWs). In particular, the tilt of GWs power spectrum (for the modes which re-enter the horizon during the reheating era) explicitly depends on the reheating EoS parameter \cite{Bernal:2019lpc,Haque:2021dha,Maity:2024odg,Barman:2023ktz,Odintsov:2024sbo} --- therefore a modification on $w_\mathrm{eff}$ in turn modifies the GWs power spectrum. From a different perspective, the reheating EoS parameter also affects the evolution of primordial magnetic field in the context of inflationary magnetogenesis, as the magnetic field's evolution through reheating stage is directly hit by $w_\mathrm{eff}$ (through Faraday's induction) \cite{Bamba:2020qdj,Maity:2021qps}. Therefore the investigation of an early universe model with the ACT data can be broadly motivated by several directions: (1) to check whether the model can trigger a viable inflationary model with respect to the ACT data, (2) to examine how the ACT data modify the post-inflationary reheating phenomenology, (3) the analysis of primordial GWs, or primordial magnetogenesis, in such model based on the ACT-DR6 data. Some of these points are being investigated in scalar-tensor inflationary models \cite{Haque:2025uri,Haque:2025uis,Chakraborty:2025oyj}.

Beside the scalar-tensor models, the higher curvature gravity models play a pivotal role in successfully describing various aspects of cosmology starting from inflation to the late time acceleration, including the reheating dynamics (see \cite{Capozziello:2011et,Nojiri:2010wj,Nojiri:2017ncd}, for an extensive review on higher curvature gravity). Theoretically, such higher curvature terms can naturally arise from the diffeomorphism property of the gravitational action. $F(R)$ gravity, Gauss-Bonnet (GB) theory of gravity or more generally Lanczos-Lovelock gravity are some of well known candidates of higher curvature theories. In general, inclusion of higher curvature terms may lead to
higher derivatives (higher than two) in the gravitational equations leading to the Ostragradsky instability. However the scalar-Einstein-GB theory is free from such instability due to the particular combination of Reimann tensor, Ricci tensor and Ricci scalar in the GB term. This is unlike to the case of $F(R)$ gravity where, in most cases, the Ostragradsky instability appears. Based on the importance of higher curvature gravity models in the context of early universe, it is utmost important to investigate (or revisit) the same from the perspective of the recent ACT data. In the present work, we examine the observational viability of non-minimally coupled scalar-Einstein-Gauss-Bonnet gravity, during inflation and post-inflationary reheating dynamics, based on the latest ACT-DR6 combined with the Planck 2018 and BAO data. Our main investigations are two folded:
\begin{itemize}
 \item Can the scalar-Einstein-GB theory of gravity trigger a viable inflation during early universe, compatible with the ACT data ?

 \item If so, then how does the ACT data influence the reheating phenomenology in the context of scalar-Einstein-GB theory ?
\end{itemize}
The scalar-Einstein-GB theory is motivated due to its rich cosmological consequences starting from early inflation to the consistency with the DESI data regarding the late dark energy era \cite{Nojiri:2005vv,Satoh:2008ck,Satoh:2007gn,Li:2007jm,Odintsov:2020sqy,Cognola:2006eg,Odintsov:2023lbb,Odintsov:2018zhw,Chakraborty:2018scm,Kanti:2015pda,Elizalde:2020zcb,Pozdeeva:2021iwc,Granda:2019jqy,Nojiri:2022xdo,Oikonomou:2024jqv,Nojiri:2023mbo,Odintsov:2025kyw,Hussain:2025vbo}. We should mention that the Gauss-Bonnet inflation in respect to the ACT data has been recently studied in \cite{Odintsov:2025wai}, however in quite different contexts. Note that, in our present analysis, we include the post-inflationary reheating dynamics to examine how the ACT data influence the reheating phenomenology as well in Gauss-Bonnet cosmology, which makes the present scenario essentially different from earlier one \cite{Odintsov:2025wai} where only the inflationary scenario is taken into account. Moreover \cite{Odintsov:2025wai} put an extra condition on the Gauss-Bonnet coupling function to make the model compatible with GW170817 event which states that the speed of gravitational waves is equal to the speed of light. However in our current paper, the model is compatible with GW170817 event, but due to the presence of reheating stage rather than incorporating any new condition on the Gauss-Bonnet coupling. Actually the speed of the gravitational wave during inflation in scalar-Einstein-GB theory is different than the speed of light (which is unity in natural units). However at the late time, in particular after the reheating phase, the scalar field energy density transforms to relativistic particles leading to the radiation dominated era with a suitable reheating temperature. In this regard, if we consider that the radiation energy does not exhibit any non-minimal coupling with the GB term, then the higher curvature GB term becomes a surface term in the four dimensional gravitational action after the reheating phase. As a consequence, the scalar-Einstein-GB gravity effectively reduces to the standard Einstein gravity with a perfect fluid (radiation) after the reheating phase, that leads to unit gravitational speed --- which is indeed consistent with GW170817 event.

\section{The model}\label{sec-model}
Let us consider the scalar-Einstein-Gauss-Bonnet gravity theory with the action:
\begin{eqnarray}
 S = \int d^4x \sqrt{-g}\left[\frac{R}{2\kappa^2} - \frac{1}{2}g^{\mu\nu}\partial_{\mu}\phi\partial_{\nu}\phi - V(\phi) + \frac{1}{2}\xi(\phi)\mathcal{G}\right]~~,
 \label{action-model}
\end{eqnarray}
where $\phi$ is the scalar field under consideration embedded having the potential $V(\phi)$, $R$ is the Ricci scalar formed by the metric $g_{\mu\nu}$ and $\kappa = \sqrt{8\pi G}$ (with $G$ being the Newton's gravitational constant). Furthermore $\mathcal{G} = R^2 - 4R_{\mu\nu}R^{\mu\nu} + R_{\mu\nu\alpha\beta}R^{\mu\nu\alpha\beta}$ represents the Gauss-Bonnet (GB) scalar and $\xi(\phi)$ is the corresponding coupling between the GB and the scalar field. Such coupling ensures the non-trivial effects of GB term in gravitational equations, otherwise the GB term has no effect for a constant $\xi$. The spatially flat FLRW metric ansatz fits our purpose in the present context, i.e
\begin{eqnarray}
 ds^2 = -dt^2 + a^2(t)\delta_{ij}dx^{i}dx^{j}~~,
 \label{metric ansatz}
\end{eqnarray}
with $t$ designates the cosmic time, $a(t)$ being the scale factor of the universe and $\frac{\dot{a}}{a} = H$ will represent the Hubble parameter 
(an overdot shows the derivative $\frac{d}{dt}$). 
Due to the spatially FLRW metric, the cosmological field equations in this context are given by,
\begin{eqnarray}
 3H^2&=&\kappa^2\left[\frac{1}{2}\dot{\phi}^2 + V(\phi) - 12\dot{\xi}H^3\right]~~,\label{FRW-1}\\
 -2\dot{H}&=&\kappa^2\left[\dot{\phi}^2 + 4\ddot{\xi}H^2 + 4\dot{\xi}H\left(2\dot{H} - H^2\right)\right]~~,\label{FRW-2}
\end{eqnarray}
and moreover, the scalar field equation turns out to be,
\begin{eqnarray}
 \ddot{\phi} + 3H\dot{\phi} + V'(\phi) - 12\xi'(\phi)\left(H^4 + H^2\dot{H}\right) = 0~~.
 \label{scalar eom}
\end{eqnarray}
Here it may be mentioned that the above three equations are not independent, actually one of them can be obtained from the other two. As mentioned earlier, we are interested on the cosmological phenomenology of the scalar-Einstein-GB model from the perspective of the ACT data, and how it influences the consecutive reheating era. For this purpose, the scalar potential and the GB coupling function are considered to have the following forms \cite{Granda:2019jqy,Odintsov:2023lbb}:
\begin{eqnarray}
 V_\mathrm{I}(\phi) = V_0\mathrm{e}^{-\lambda\kappa\phi}~~~~~\mathrm{and}~~~~~\xi_\mathrm{I}(\phi) = \xi_0\mathrm{e}^{-\eta\kappa\phi}~,\label{inf-pot}
\end{eqnarray}
during inflation, and
\begin{eqnarray}
 V_\mathrm{R}(\phi) = V_1\mathrm{e}^{2\left(\phi-\phi_\mathrm{s}\right)/\phi_\mathrm{0}}
 ~~~~\mathrm{and}~~~~\xi_\mathrm{R}(\phi) = \xi_1\mathrm{e}^{-2\left(\phi - \phi_\mathrm{s}\right)/\phi_\mathrm{0}}~,\label{reh-pot}
\end{eqnarray}
during reheating era. Here both $V_0$ and $V_1$ have mass dimension [+4], while $\xi_0$ and $\xi_1$ are dimensionless parameters. The motivation for such form of potentials are as follows: (1) the inflationary potentials are able to produce a viable quasi de-Sitter inflation, as we will see; (2) the reheating $V_\mathrm{R}(\phi)$ and
$\xi_\mathrm{R}(\phi)$ in Eq.~(\ref{reh-pot}) are motivated as they lead to a power law form of Hubble parameter which are suitable for the universe's evolution after inflation; and moreover, (3) the respective forms of $V(\phi)$ and $\xi(\phi)$ are joined smoothly at the junction between inflation-to-reheating, as shown in \cite{Odintsov:2023lbb}.

\section{Inflationary era}\label{sec-inf}
The slow roll parameters during inflation, for the Einstein-GB theory, are given by \cite{Guo:2010jr},
\begin{eqnarray}
 \epsilon_0&=&-\dot{H}/H^2~~~~~~~~~~,~~~~~~~~~~~\epsilon_1 = \frac{\dot{\epsilon}_0}{H\epsilon_0}~~,\nonumber\\
 \delta_0&=&-4\dot{\xi}H~~~~~~~~~~~~,~~~~~~~~~~~~\delta_1 = \frac{\dot{\delta}_0}{H\delta_0}~.
 \label{Sr quantities}
\end{eqnarray}
The slow-roll parameters are written in the unit of $\kappa = 1$; in fact, all the subsequent calculations are based in this unit. It is noted that the first two slow roll parameters are similar to that of in the usual scalar-tensor theory, while the last two parameters ($\delta_0$ and $\delta_1$) arise for the modified Einstein-GB theory under consideration. With the slow roll conditions, given by $\epsilon_{i}\ll1$ and $\delta_{i} \ll 1$, the cosmological field equations turn out to be,
\begin{eqnarray}
 0&=&-3H^2 + V_\mathrm{I}(\phi)~~,\nonumber\\
 0&=&2\dot{H} + \left[\dot{\phi}^2 - 4\dot{\xi}_\mathrm{I}H^3\right]~~,\nonumber\\
 0&=&3H\dot{\phi} + V_\mathrm{I}'(\phi) - 12\xi_\mathrm{I}'(\phi)H^4~~.
 \label{SR eom}
\end{eqnarray}
Owing to these equations, the slow roll parameters, in terms of $V_\mathrm{I}(\phi) = V_0\mathrm{e}^{-\lambda\phi}$ and
$\xi_\mathrm{I}(\phi) = \xi_0\mathrm{e}^{-\eta\phi}$, turn out to be,
\begin{eqnarray}
 \epsilon_0&=&\frac{\lambda}{6}\left\{3\lambda - 4V_0\xi_0\eta\mathrm{e}^{-\left(\lambda + \eta\right)\phi}\right\}~~,\nonumber\\
 \delta_0&=&\frac{4}{9}V_0\xi_0\eta\left\{3\lambda\mathrm{e}^{\left(\lambda + \eta\right)\phi} - 4V_0\xi_0\eta\right\}
 \mathrm{e}^{-2\left(\lambda + \eta\right)\phi}~.\label{SR-1}
\end{eqnarray}
and
\begin{eqnarray}
 \epsilon_1&=&\frac{4}{3}V_0\xi_0\eta\left(\lambda + \eta\right)\mathrm{e}^{-\left(\lambda + \eta\right)\phi}~~,\nonumber\\
 \delta_1&=&-\frac{1}{3}\left(\lambda + \eta\right)\left\{3\lambda - 8V_0\xi_0\eta\mathrm{e}^{-\left(\lambda + \eta\right)\phi}\right\}~~.
 \label{SR-2}
\end{eqnarray}
respectively, where the overprime with the argument $\phi$ symbolizes $\frac{d}{d\phi}$. Let us consider that the scalar field, at the end of inflation, is given by $\phi = \phi_\mathrm{f}$ (the subscript 'f' denotes at the end of inflation), then, Eq.~(\ref{SR-1}) gives the following condition on $\phi_\mathrm{f}$:
\begin{eqnarray}
 \mathrm{e}^{2\left(\lambda + \eta\right)\phi_\mathrm{f}} + \frac{4}{3}V_0\xi_0\eta\lambda\mathrm{e}^{\left(\lambda + \eta\right)\phi_\mathrm{f}} 
 - \frac{16}{9}\left(V_0\xi_0\eta\right)^2 = 0\nonumber
\end{eqnarray}
which can be solved for $\phi_\mathrm{f}$ to get,
\begin{eqnarray}
 \phi_\mathrm{f} = \frac{1}{\left(\lambda + \eta\right)}\ln{\left[\frac{2}{3}V_0\xi_0\eta\left\{\sqrt{\left(\lambda^2 + 4\right)} - \lambda\right\}\right]}~~.
 \label{end phi}
 \end{eqnarray}
The above expression leads to the inflationary e-folding number ($N_\mathrm{f}$), and is given by:
 \begin{eqnarray}
  N_\mathrm{f} = \frac{1}{\lambda\left(\lambda + \eta\right)}
  \ln{\left[\frac{3\lambda\mathrm{e}^{\left(\lambda + \eta\right)\phi_\mathrm{f}} - 4V_0\xi_0\eta}
  {3\lambda\mathrm{e}^{\left(\lambda + \eta\right)\phi_\mathrm{i}} - 4V_0\xi_0\eta}\right]}~~,
  \label{e-fold-2}
 \end{eqnarray}
 where $\phi_\mathrm{i}$ is the scalar field at the beginning of inflation ($N=0$). In the present work, the beginning of inflation is taken to be the instance when the CMB scale mode ($\sim 0.005 \mathrm{Mpc}^{-1}$) crosses the horizon. The $\phi_\mathrm{i}$ can be obtained by inverting Eq.~(\ref{e-fold-2}), as follows:
 \begin{eqnarray}
  \phi_\mathrm{i} = \frac{1}{\left(\lambda + \eta\right)}
  \ln{\left[\frac{V_0\xi_0\eta}{3\lambda}\left\{2\mathrm{e}^{-\lambda\left(\lambda + \eta\right)N_\mathrm{f}}\left(\lambda\sqrt{\left(\lambda^2 + 4\right)} - \lambda^2 - 2\right) + 4\right\}\right]}~~,
  \label{start phi}
 \end{eqnarray}
 where we use the $\phi_\mathrm{f}$ from Eq.~(\ref{end phi}). Moreover, Eq.~(\ref{start phi}) and Eq.~(\ref{end phi}) result to the evolution of the scalar field during inflation (in terms of e-fold number), i.e. $\phi = \phi(N)$, as,
 \begin{eqnarray}
  \phi(N) = \frac{1}{\left(\lambda + \eta\right)}
  \ln{\left[\frac{V_0\xi_0\eta}{3\lambda}\left\{2\mathrm{e}^{-\lambda\left(\lambda + \eta\right)\left(N_\mathrm{f} - N\right)}
  \left(\lambda\sqrt{\left(\lambda^2 + 4\right)} - \lambda^2 - 2\right) + 4\right\}\right]}~~.
  \label{e-fold-4}
 \end{eqnarray}
 Due to this form of $\phi(N)$, Eq.~(\ref{SR eom}) immediately leads to $H = H(N)$ as,
 \begin{eqnarray}
  H(N) = \left(\frac{V_0}{3}\right)^{\frac{1}{2}}
  \left[\frac{V_0\xi_0\eta}{3\lambda}\left\{2\mathrm{e}^{-\lambda\left(\lambda + \eta\right)\left(N_\mathrm{f} - N\right)}
  \left(\lambda\sqrt{\left(\lambda^2 + 4\right)} - \lambda^2 - 2\right) + 4\right\}\right]^{\frac{-\lambda}{2\left(\lambda + \eta\right)}}~~.
  \label{Hubble-evolution}
 \end{eqnarray}
Thus as a whole, Eq.~(\ref{e-fold-4}) and Eq.~(\ref{Hubble-evolution}) represent the evolution of the scalar field and of the Hubble parameter during inflation respectively. Having set the stage, our main aim is to investigate the compatibility of the current model with the ACT data \cite{ACT:2025fju,ACT:2025tim}. For this purpose, we need to go through the (scalar and tensor) cosmological perturbation started from the Bunch-Davies vacuum.

\subsection*{Cosmological phenomenology}

\begin{figure}[!h]
\begin{center}
\centering
\includegraphics[width=3.5in,height=2.5in]{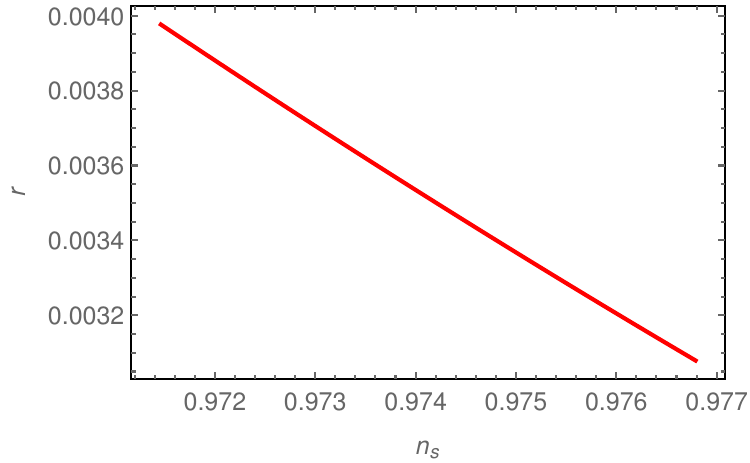}
\caption{Parametric plot of $n_s$ vs. $r$ with $\lambda = -0.008$ and $\eta = 1$, in which case, the $n_s$ and $r$ get simultaneously compatible with the Panck data for $N_\mathrm{f} = [55,65]$.}
\label{plot-observable}
\end{center}
\end{figure}

The scalar and tensor perturbation over the spatially flat FRW background metric (\ref{metric ansatz}) are defined as:
\begin{eqnarray}
 ds^2 = -\left(1 + 2\psi\right)dt^2 + a^2(t)\left(1 - 2\psi\right)\delta_{ij}dx^{i}dx^{j}~~,\label{sp}
\end{eqnarray}
and
\begin{eqnarray}
 ds^2 = -dt^2 + a^2(t)\left(\delta_{ij} + h_\mathrm{ij}\right)dx^{i}dx^{j}~~,\label{tp}
\end{eqnarray}
where $\psi(t,\vec{x})$ and $h_\mathrm{ij}(t,\vec{x})$ represent the scalar and tensor perturbation variable respectively. In the comoving gauge where the velocity perturbation is taken to be zero, the scalar perturbation $\psi$ in Eq.~(\ref{sp}) resembles to the curvature perturbation, i.e $\mathcal{R} = \psi$. Thus we can work with the variable $\psi(t,\vec{x})$. The Fourier modes of the curvature perturbation and the tensor perturbation, at linear order in perturbation theory, satisfy the respective Mukhanov-Sasaki equations:
\begin{eqnarray}
 v_\mathrm{s}''(k,\tau) + \left(c_\mathrm{s}^2k^2 - \frac{z_\mathrm{s}''}{z_\mathrm{s}}\right)v_\mathrm{s}(k,\tau) = 0~~,
 \label{cp-1}
\end{eqnarray}
and
\begin{eqnarray}
 v_\mathrm{T}''(k,\tau) + \left(c_\mathrm{T}^2k^2 - \frac{z_\mathrm{T}''}{z_\mathrm{T}}\right)v_\mathrm{T}(k,\tau) = 0~~,
 \label{tp-2}
\end{eqnarray}
respectively. Here $v_\mathrm{s}(k,\tau) = z_\mathrm{s}\mathcal{R}$ and $v_\mathrm{T}(k,\tau) = z_\mathrm{T}(\tau)h(k,\tau)$ are the Mukhanov-Sasaki variables for the curvature perturbation and for the tensor perturbation respectively, $k$ is the momentum of the perturbation mode and $\tau$ is the conformal time defined by $d\tau = \frac{dt}{a(t)}$. The functions mentioned in the perturbation equations have the following forms, in the present context of scalar-Einstein-GB theory \cite{Hwang:2005hb,Noh:2001ia}:
\begin{eqnarray}
 z_\mathrm{s}^2 = a^2\left[\frac{\dot{\phi}^2 - 6\Delta\dot{\xi}H^3}{H^2\left(1 - \frac{\Delta}{2}\right)^2}\right]~~~~~~~~~~~~,~~~~~~~~~~~
 c_\mathrm{s}^2 = 1 - \frac{8\Delta\dot{\xi}H\dot{H} + 2\Delta^2H^2\left(\ddot{\xi} - \dot{\xi}H\right)}{\dot{\phi}^2 - 6\Delta\dot{\xi}H^3}~~,
 \label{cp-3}
\end{eqnarray}
and
\begin{eqnarray}
 z_\mathrm{T}^2 = a^2\left\{1 + 4\dot{\xi}H\right\}~~~~~~,~~~~~c_\mathrm{T}^2 = 1 + \frac{4\left(\ddot{\xi} - \dot{\xi}H\right)}{1 + 4\dot{\xi}H}~~,
 \label{tp-3}
\end{eqnarray}
where $\Delta = \frac{\delta_0}{1-\delta_0}$ (recall that $\delta_0 = -4\dot{\xi}H$ is the slow roll parameter). By using Eq.~(\ref{SR-1}) and Eq.~(\ref{SR-2}), we determine the following ingredients necessary to solve the perturbation evolutions, namely,
\begin{eqnarray}
 \frac{z_\mathrm{s}''}{z_\mathrm{s}}&=&\frac{1}{\tau^2}\left\{2 + 3\epsilon_0 + \frac{3}{2}\left(\frac{2\epsilon_0\epsilon_1 - \delta_0\delta_1}{2\epsilon_0 - \delta_0}\right)\right\}~~,\nonumber\\
 c_\mathrm{s}^2&=&1 - \Delta^2\left\{\frac{2\epsilon_0 + \frac{\delta_0}{2}\left(1 - 5\epsilon_0 - \delta_1\right)}{2\epsilon_0 - \delta_0 + \frac{3}{2}\Delta\delta_0}\right\}~~,
 \label{cp-8}
\end{eqnarray}
and
\begin{eqnarray}
 \frac{z_\mathrm{T}''}{z_\mathrm{T}}&=&\frac{1}{\tau^2}\bigg\{(2 + 3\epsilon_0)\bigg\}~~,\nonumber\\
 c_\mathrm{T}^2&=&(1 + \delta_0) + \mathrm{H.O.~in~slow~roll~parameters}~~,
 \label{tp-7}
\end{eqnarray}
by keeping the leading order term of the slow roll parameters. With such expressions, the curvature and the tensor perturbations power spectra at super-Hubble regime comes as,
\begin{eqnarray}
 \mathcal{P}_\mathrm{s}(k,\tau)&=&\left(\frac{H^2}{4\pi^2}\right)\left(\frac{1}{aH|\tau|}\right)^2\left(\frac{1}{c_\mathrm{s}^3\left(2\epsilon_0 - \delta_0\right)}\right)\frac{\Gamma^2(\nu_s)}{\Gamma^2(3/2)}\left(\frac{c_\mathrm{s}k|\tau|}{2}\right)^{3-2\nu_s}~~,\nonumber\\
 \mathcal{P}_\mathrm{T}(k,\tau)&=&\left(\frac{H^2}{4\pi^2}\right)\left(\frac{1}{aH|\tau|}\right)^2\left(\frac{8}{c_\mathrm{T}^3\left(1 - \delta_0\right)}\right)\frac{\Gamma^2(\nu_T)}{\Gamma^2(3/2)}\left(\frac{c_\mathrm{T}k|\tau|}{2}\right)^{3-2\nu_T}~~,
 \label{tp-9}
\end{eqnarray}
where the exponents $\nu_s$ and $\nu_T$ are of the forms as,
\begin{eqnarray}
 \nu_s^2&=&\frac{1}{4} + \left\{2 + 3\epsilon_0 + \frac{3}{2}\left(\frac{2\epsilon_0\epsilon_1 - \delta_0\delta_1}{2\epsilon_0 - \delta_0}\right)\right\}~~,\nonumber\\
\nu_T^2&=&\frac{1}{4} + \left(2 + 3\epsilon_0\right)~~.
\nonumber
 \end{eqnarray}
 Such primordial power spectra immediately lead to observable quantities like the spectral tilt for curvature perturbation ($n_s$) and the tensor-to-scalar ratio ($r$) which are defined by,
\begin{eqnarray}
 n_s = \frac{\partial\ln{\mathcal{P}_\mathrm{s}}}{\partial\ln{k}}\bigg|_{h.c}~~~~~~~~~~~~~~~~\mathrm{and}~~~~~~~~~~~r = \frac{\mathcal{P}_\mathrm{T}}{\mathcal{P}_\mathrm{s}}\bigg|_{h.c}~~,
 \label{ct-1}
\end{eqnarray}
where the suffix 'h.c' indicates the horizon crossing instance of the CMB scale mode on which we are interested to evaluate the observable quantities (the horizon crossing of $k$-th mode occurs at $k = aH$ or $|\tau| = 1/k$). Due to Eq.(\ref{tp-9}), the $n_s$ and $r$ in the present context comes with the following forms:
\begin{eqnarray}
 n_s = 4 - 2\nu_s~~~~~~~~\mathrm{and}~~~~~~~~r = \frac{8\left(2\epsilon_0 - \delta_0\right)}{\left(1 - \delta_0\right)}\left(\frac{\left(c_\mathrm{s}/2\right)^{2\nu_s}}{\left(c_\mathrm{T}/2\right)^{2\nu_T}}\right)\frac{\Gamma^2(\nu_T)}{\Gamma^2(\nu_s)}~~,
 \label{ct-2}
\end{eqnarray}
where $c_\mathrm{s}$, $c_\mathrm{T}$, $\nu_s$ and $\nu_T$ are shown above. Using such, the final forms of $n_s$ and $r$ (in terms of model parameters) come as,
\begin{eqnarray}
 n_s&=&1 - \lambda^2 - \frac{2\left(\lambda + 2\eta\right)~\mathrm{e}^{\lambda\left(\lambda + \eta\right)N_\mathrm{f}}}
 {\lambda\sqrt{\lambda^2 + 4} - \left(\lambda^2 + 2 - 2\mathrm{e}^{\lambda\left(\lambda + \eta\right)N_\mathrm{f}}\right)}~~,\nonumber\\
 r&=&8\lambda^2\left[\frac{\lambda\sqrt{\lambda^2 + 4} - \left(\lambda^2 + 2\right)}
 {\lambda\sqrt{\lambda^2 + 4} - \left(\lambda^2 + 2 - 2\mathrm{e}^{\lambda\left(\lambda + \eta\right)N_\mathrm{f}}\right)}\right]^2
 \label{observable-3}
\end{eqnarray}
respectively. We now examine the viability of the model with the recent ACT-DR6, combined with the Planck and BAO, data which puts a constraint on $n_s$ and $r$ by \cite{ACT:2025fju,ACT:2025tim},
\begin{eqnarray}
 n_s = 0.9743 \pm 0.0034~~~~~~~~\mathrm{and}~~~~~~~~r < 0.038~~.\nonumber
\end{eqnarray}
Eq.(\ref{observable-3}) clearly reveals that that the theoretical expectations of $n_s$ and $r$ depend on the model parameters $\lambda$ and $\eta$, and on the inflationary e-fold number $N_\mathrm{f}$. The above expressions of $n_s$ and $r$ turn out to be simultaneously compatible with the ACT-DR6+Planck18+BAO data for suitable values of the model parameters -- these are depicted in the parametric plots of Fig.~[\ref{plot-observable}]. This shows that the theoretical expectations of $n_s$ and $r$ get consistent with the ACT+Planck+BAO data provided the inflationary e-fold lies within $[55,65]$, i.e $N_\mathrm{f} = [55,65]$, with $\lambda = -0.008$ and $\eta = 1$.

Therefore the scalar-Einstein-GB theory, with $V_\mathrm{I}(\phi) = V_0\mathrm{e}^{-\lambda\phi}$ and $\xi_\mathrm{I}(\phi) = \xi_0\mathrm{e}^{\eta\phi}$, is able to trigger a quasi de-Sitter inflationary scenario during the early universe that is compatible with the recent ACT-DR6 data (combined with the Planck and BAO) and also has an exit at a finite e-fold given by $N_\mathrm{f} = [55,65]$. At this stage, it is important to examine whether such viable ranges of the model parameters (with respect to the ACT-DR6 data) is compatible with the reheating era, or, how the parameter spaces get more constrained due to the post-inflationary reheating era.\\

On a different perspective, here it may be mentioned that the author of \cite{Pozdeeva:2020shl} considered a different set of scalar potential and GB coupling function in order to examine the attractor behaviour along with observational viability (during inflation) of scalar-Einstein-GB theory. In particular, \cite{Pozdeeva:2020shl} used the following sets of $V(\phi)$ and $\xi(\phi)$:
\begin{eqnarray}
 V(\phi) \sim \mathrm{exp}\left[-2\mathrm{e}^{-p \phi}\right]~~~~~~~\mathrm{and}~~~~~~\xi(\phi) \sim \mathrm{exp}\left[2\mathrm{e}^{-p \phi}\right] \, ,
 \label{n2}
 \end{eqnarray}
where $p$ is a constant. For such $V(\phi)$ (and $\xi(\phi)$), the inflationary parameters (like $n_\mathrm{s}$ and $r$) in the large field regime depends on the factor $\mathrm{e}^{-p\phi}$ \cite{Pozdeeva:2020shl}. As a result, it turns out that with the condition: $\mathrm{e}^{-p \phi} \ll 1$, the above scalar-Einstein-GB model (in Eq.~(\ref{n2})) becomes a viable one.

\section{Influence of ACT-DR6 on reheating era}\label{sec-reh}
The universe enters to reheating era after the inflaytion ends, during which, the scalar field potential and the GB coupling function in the present context follow Eq.(\ref{reh-pot}). In the context of scalar-Einstein-GB theory, the scalar field energy density is different (explicitly) than the usual scalar-tensor theory. In particular, the effective energy density and the effective pressure of the scalar field in present model are given by:
\begin{eqnarray}
 \rho_\mathrm{\phi} = \frac{1}{2}\dot{\phi}^2 + V(\phi) - 12\dot{\xi}H^3~~,
 \label{reh-1}
\end{eqnarray}
and
\begin{eqnarray}
 p_\mathrm{\phi} = \dot{\phi}^2 + 4\ddot{\xi}H^2 + 4\dot{\xi}H\left(2\dot{H} - H^2\right) - \rho_\mathrm{\phi}~~,
 \label{reh-2}
\end{eqnarray}
respectively. Therefore the effective equation of state (EoS) parameter during the reheating stage follows:
\begin{eqnarray}
 w_\mathrm{eff} = \frac{p_\mathrm{\phi}}{\rho_\mathrm{\phi}} = -1 + \frac{\dot{\phi}^2 + 4\ddot{\xi}H^2 + 4\dot{\xi}H\left(2\dot{H} - H^2\right)}
 {\frac{1}{2}\dot{\phi}^2 + V(\phi) - 12\dot{\xi}H^3}~~.
 \label{reh-3}
\end{eqnarray}
During the reheating era, the scalar field energy density ($\rho_\mathrm{\phi}$) decays to relativistic particles with a certain decay width (symbolized by $\Gamma$). The Hubble parameter during the reheating stage is generally considered to have a power law form. This is due to the fact that the effective decay of the scalar field to relativistic particles occur around the end of reheating when the Hubble parameter becomes comparable to the decay width. Thus we take the following power law ansatz :
\begin{eqnarray}
 H(t) = m/t~~,
 \label{reh-4}
\end{eqnarray}
during the reheating era, where $m$ is the corresponding exponent. With $H(t) = \frac{m}{t}$ along with the scalar potential and the GB coupling function during the reheating era (shown in Eq.(\ref{reh-pot})), the field equations (see Eq.(\ref{FRW-1}), Eq.(\ref{FRW-2}) and Eq.(\ref{scalar eom})) immediately lead to the scalar field's evolution during the reheating phase as,
\begin{eqnarray}
 \phi(t) = \phi_\mathrm{s} - \phi_\mathrm{0}~\mathrm{ln}{t}~~,
 \label{reh-6}
\end{eqnarray}
provided $\phi_\mathrm{0}$ is connected with $V_\mathrm{1}$ and $\xi_\mathrm{1}$ by the following way:
\begin{eqnarray}
 V_\mathrm{1}&=&\frac{\left(5m^2 - m\right)}{(1 + m)}\left(\frac{\phi_\mathrm{0}^2}{2m} - 1\right) + 3m^2 - m~~,\nonumber\\
 \xi_\mathrm{1}&=&\frac{1}{4m(1 + m)}\left(\frac{\phi_\mathrm{0}^2}{2m} - 1\right)~~.
 \label{reh-7}
\end{eqnarray}
For having simpler expressions, we denote $\frac{1}{(1 + m)}\left(\frac{\phi_\mathrm{0}^2}{2m} - 1\right) = c_1$ (say) from onwards. Thus as a whole, the Hubble parameter and the scalar field, during the reheating era, follow Eq.(\ref{reh-4}) and Eq.(\ref{reh-6}) respectively where $\phi_\mathrm{0}$ being connected with $V_\mathrm{1}$ and $\xi_\mathrm{1}$ by Eq.(\ref{reh-7}). The Hubble evolution from Eq.~(\ref{reh-4}) immediately leads to the reheating EoS parameter as,
\begin{eqnarray}
 w_\mathrm{eff} = -1 - \frac{2\dot{H}}{3H^2} = -1 + \frac{2}{3m}
 \label{reh-5}
\end{eqnarray}
which is actually a constant, as expected.

The reheating era is mainly parametrized by: $\left\{w_\mathrm{eff}, N_\mathrm{re}, T_\mathrm{re}\right\}$ where $N_\mathrm{re}$ and $T_\mathrm{re}$ represent the e-fold number and the temperature of the reheating era. Based on $H = \frac{m}{t}$ or equivalently $H \propto a^{-\frac{3}{2}\left(1 + w_\mathrm{eff}\right)}$ (where $w_\mathrm{eff} = -1 + \frac{2}{3m}$ and $a$ is the scale factor of the universe), we have the following expressions of $N_\mathrm{re}$ and $T_\mathrm{re}$ as follows \cite{Odintsov:2023lbb}:
\begin{eqnarray}
 N_\mathrm{re} = \frac{4}{\left(1 - 3w_\mathrm{eff}\right)}\left\{61.6 - \ln{\left[\frac{\left(3H_\mathrm{f}^2\right)^{1/4}}{H_\mathrm{i}}\right]} - N_\mathrm{f}\right\}~~.
 \label{et-7}
\end{eqnarray}
and
\begin{eqnarray}
 T_\mathrm{re} = H_\mathrm{i}\left(\frac{43}{11g_\mathrm{re}}\right)^{1/3}\left(\frac{T_0}{k/a_0}\right)e^{-\left(N_\mathrm{f} + N_\mathrm{re}\right)}~~,
 \label{et-4}
\end{eqnarray}
respectively, where the subscript '0' and 're' denote the present epoch and the end of reheating, respectively, with $T_0 = 2.93\mathrm{K}$. Moreover $k$ is the large scale (or CMB scale) mode given by: $\frac{k}{a_0} \sim 0.05\mathrm{Mpc}^{-1}$. In the above expression, $g_\mathrm{re}$ is the relativistic degrees of freedom (generally taken to be $\approx 100$). Furthermore the $H_\mathrm{i}$ and $H_\mathrm{f}$ in the above equations represent the Hubble parameter at the beginning and at the of inflation respectively, which have the following forms (in terms of the model parameters):
\begin{eqnarray}
 H_\mathrm{i}^2 = \left(\frac{V_0}{3}\right)e^{-\lambda\phi_\mathrm{i}}~~,
 \label{et-8}
\end{eqnarray}
and
\begin{eqnarray}
 H_\mathrm{f}^2 = \left(\frac{V_0}{6}\right)e^{-\lambda\phi_\mathrm{f}}~~,
 \label{et-9}
\end{eqnarray}
with $\phi_\mathrm{i}$ and $\phi_\mathrm{f}$ are shown in Eq.~(\ref{start phi}) and in Eq.(\ref{end phi}). Owing to the above forms of of $H_\mathrm{i}$ and $H_\mathrm{f}$, the $N_\mathrm{re}$ from Eq.(\ref{et-7}) comes as (in terms of model parameters),
\begin{eqnarray}
 N_\mathrm{re} = \frac{4}{\left(1 - 3w_\mathrm{eff}\right)}\left\{61.6 - \frac{1}{4}\ln{\left(\frac{9e^{\lambda\left(2\phi_\mathrm{i} - \phi_\mathrm{f}\right)}}{2V_0}\right)} - N_\mathrm{f}\right\}~~.
 \label{et-10}
\end{eqnarray}
Moreover, Eq.~(\ref{et-4}) provides the reheating temperature as,
\begin{eqnarray}
 T_\mathrm{re} = \left(\frac{V_0e^{-\lambda\phi_\mathrm{i}}}{3}\right)^{1/2}\left(\frac{43}{11g_\mathrm{re}}\right)^{1/3}\left(\frac{T_0}{k/a_0}\right)
 \mathrm{exp}\left[\frac{3\left(1 + w_\mathrm{eff}\right)}{\left(1 - 3w_\mathrm{eff}\right)}N_\mathrm{f} - \frac{4}{\left(1 - 3w_\mathrm{eff}\right)}\left\{61.6 - \frac{1}{4}\ln{\left(\frac{9e^{\lambda\left(2\phi_\mathrm{i} - \phi_\mathrm{f}\right)}}{2V_0}\right)}\right\}\right]~~.
 \label{et-11}
\end{eqnarray}

\begin{figure}[!h]
\begin{center}
\centering
\includegraphics[width=3.0in,height=2.0in]{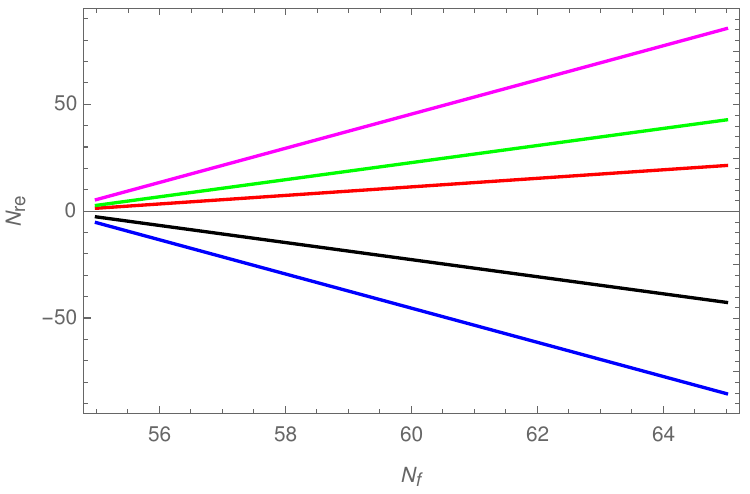}
\includegraphics[width=3.0in,height=2.0in]{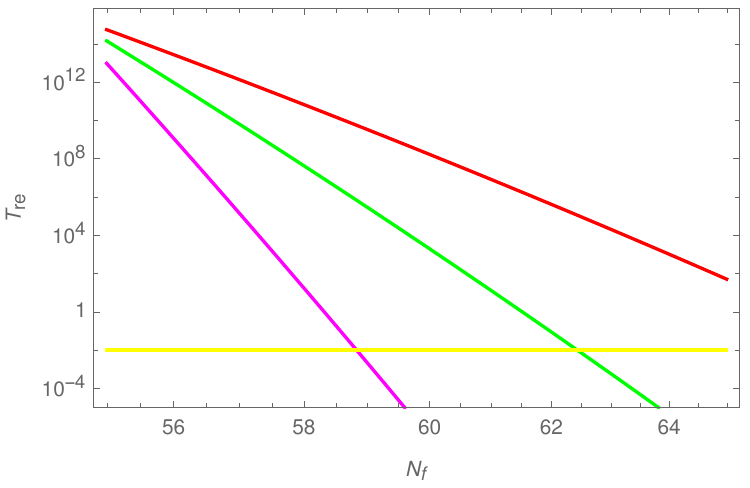}
\caption{{\color{blue}Left Plot}: $N_\mathrm{re}$ vs. $N_\mathrm{f}$; {\color{blue}Right Plot}: $T_\mathrm{re}$ (in GeV unit) vs. $N_\mathrm{f}$ for various values of $w_\mathrm{eff}$. In both the plots, we consider $\lambda = -0.005$ $\eta = 1$ and $V_0 = 10^{-12}$ (in Planck units). The reheating EoS parameter is taken as $w_\mathrm{eff} = 0 ~(\mathrm{Black~curve}), \frac{1}{6}~(\mathrm{Blue~curve}), \frac{1}{2}~(\mathrm{Magenta~curve}), \frac{2}{3}~(\mathrm{Green~curve}), 1~(\mathrm{Red~curve})$ respectively. Moreover the yellow curve in the right plot is the BBN temperature $\sim 10^{-2}\mathrm{GeV}$. In both the plots, we take $\lambda = -0.008$ and $\eta = 1$, same as in the inflationary case.}
\label{plot-R1}
\end{center}
\end{figure}

Recall that the model parameters get constrained from inflationary phenomenology as in Fig.~[\ref{plot-observable}]. We now examine whether such constraints on the model parameters coming from the inflationary phenomenology are supported or ruled out or get further constrained by the reheating stage. For this purpose, we give the plots for $N_\mathrm{re}$ vs. $N_\mathrm{f}$ and $T_\mathrm{re}$ vs. $N_\mathrm{f}$ for various values of $w_\mathrm{eff}$. In particular, we take $w_\mathrm{eff} = 0, \frac{1}{6}, \frac{1}{2}, \frac{2}{3}, 1$ respectively. Fig.[\ref{plot-R1}] clearly demonstrates that $N_\mathrm{f} = [55,65]$ allows only $w_\mathrm{eff} > \frac{1}{3}$ (i.e. such reheating EoS parameter which is larger than the value $\frac{1}{3}$), otherwise the reheating e-fold number becomes negative and thus is unphysical. Taking such constraint on $w_\mathrm{eff}$ into account, we plot $T_\mathrm{re}$ vs. $N_\mathrm{f}$ for the aforementioned values of $w_\mathrm{eff}$, which depicts that the $T_\mathrm{re}$ remains less than the BBN temperature ($T_\mathrm{BBN} \sim 10^{-2}\mathrm{GeV}$) for $w_\mathrm{eff} > 1/3$. Importantly, the plot of $T_\mathrm{re}$ vs. $N_\mathrm{f}$ demonstrates that each allowed value of $\frac{1}{3} < w_\mathrm{eff} < 1$ puts an individual constraint on the inflationary e-fold number in order to satisfy $T_\mathrm{re} \gtrsim T_\mathrm{BBN}$ --- for instance, $w_\mathrm{eff} = \frac{1}{2}$ is consistent with $N_\mathrm{f} = [55,58.5]$.\\

Thus combining the inflation(ACT-DR6+Planck+BAO) and reheating constraints, what we find is demonstrated in the Table.~[\ref{Table-0}]:

\begin{table}[h]
  \centering
 {%
  \begin{tabular}{|c|c|c|}
   \hline
    Model parameters & Reheating EoS parameter & Constraint on $N_\mathrm{f}$ and $N_\mathrm{re}$\\

   \hline
  \hline\\

   $\lambda = -0.005$, $\eta = 1$ & (a) $w_\mathrm{eff} < \frac{1}{3}$ & $N_\mathrm{re} < 0$ and thus ruled out\\

   \hline\\

     & (b) $\frac{1}{3} < w_\mathrm{eff} < 1$ & $N_\mathrm{re} > 0$ and $N_\mathrm{f} = [55,65]$ depending on the specific value of $w_\mathrm{eff}$\\

   \hline
   \hline
  \end{tabular}%
 }
  \caption{Constraints coming from both the inflation (ACT-DR6+Planck+BAO) and reheating phenomenology).}
  \label{Table-0}
 \end{table}

Thus we may notice that the viable range inflationary e-fold number coming from ACT-DR6+Planck+BAO data gets further constrained by the input of the reheating stage. Such constraints are different than the case where only Planck data is considered, see \cite{Odintsov:2023lbb}. In particular, Table.~[\ref{Table-0}] shows that $w_\mathrm{eff} < 1/3$ is ruled out from the perspective of ACT-DR6 data, which is unlike to the case where only Planck data is considered that indeed allows $w_\mathrm{eff} < 1/3$ in the context of scalar-Einstein-GB cosmological scenario. This indicates that the inflationary ACT-DR6 data substantially influences the post-inflationary reheating stage.

\section{Conclusion}
The work examines the observational viability of a higher curvature gravity, namely the non-minimally coupled scalar-Einstein-GB theory, from the perspective of the recent ACT-DR6+Planck18+BAO data. The scalar-Einstein-GB theory leads to Ostragradsky free model as well as it is also compatible with the GW170817 event at late time. Our analysis not only provides the updated constraints of inflationary phenomenology, but also extends up to the reheating dynamics, in particular, we examine how the ACT data influences the reheating phenomenology in the context of scalar-Einstein-GB cosmology. It turns out that, with exponential forms of scalar potential and the non-minimal GB coupling function, the present model is able to trigger a viable inflation during early universe, that is well compatible with the ACT-DR6+Planck18+BAO data. The updated constraints on various parameters are verified, and it seems that the ACT data considerably affects the inflationary e-fold number compared to the case where only Planck 2018 is taken into account. With such constraints coming from the inflationary phenomenology, we address the post-inflation reheating dynamics in the present context --- this shows that the scalar-Einstein-GB theory allows a power law solution of the HUbble parameter during the reheating stage, and thus, it is parametrized by a constant reheating EoS parameter (symbolized by $w_\mathrm{eff}$). Regarding the reheating phenomenology, our key findings are as follows: (1) the ACT data prefers only $w_\mathrm{eff} > 1/3$, otherwise the reheating e-fold number becomes negative (within the allowed range of inflationary e-fold number) which is unphysical; (2) we examine the reheating temperature ($T_\mathrm{re}$) with respect to the inflationary e-fold for various values of $\frac{1}{3} < w_\mathrm{eff} < 1$, which shows that each allowed value of $w_\mathrm{eff}$ puts an individual constraint on the inflationary e-fold number in order to satisfy $T_\mathrm{re} \gtrsim T_\mathrm{BBN}$. These are unlike to that of when only Planck 2018 is considered --- this indicates that, beside the inflationary phenomenology, the ACT data also substantially affects the reheating era in the context of scalar-Einstein-GB theory. The allowed range of $\frac{1}{3} < w_\mathrm{eff} < 1$ in scalar-Einstein-GB cosmology is different than the well known $\alpha$-attractor inflationary models where $w_\mathrm{eff} < 1/3$ is also allowed even with the recent ACT data \cite{Haque:2025uri}.

The updated viable range of $w_\mathrm{eff}$, based on the ACT data, in turn modifies the primordial gravitational waves (GWs) spectrum for the modes that re-enter the horizon during the reheating stage. In particular, any value of $w_\mathrm{eff} > 1/3$ is expected to enhance the primordial GWs amplitude (for the reheating-modes) and allows to cross the sensitivity curves of various GWs observatories. The primordial gravitational waves in higher curvature cosmology with the ACT data is an important aspect and is worthwhile to investigate in future.

\bibliography{bibliography}
\bibliographystyle{./utphys1}

\end{document}